\documentclass[12pt,letterpaper]{article}

\usepackage[includeheadfoot,  
            marginratio={1:1,2:3},  
            width=433pt,  
            height=688pt,]{geometry} 
            
\usepackage{times}
\usepackage{amsmath}
\usepackage{amssymb}
\usepackage[]{graphicx}
\DeclareMathOperator{\sign}{sign}
\newcommand{\ov}{\overline}

\begin{document}

\vspace*{-1.5cm} 
\begin{flushright} 
  {\small 
  MPP-2007-187 \\ 
  LMU-ASC 74/07\\ 
  } 
\end{flushright}

\vspace{1.5cm} 
\begin{center} 
  {\LARGE 
  D-brane Instantons in\medskip\\ 4D Supersymmetric String Vacua 
  } 
\end{center}

\vspace{0.25cm} 
\begin{center} 
  {\small 
  Nikolas Akerblom$^1$, Ralph~Blumenhagen$^1$, Dieter L\"ust$^{1,2}$, \\ 
  Maximilian Schmidt-Sommerfeld$^1$ \\ 
  } 
\end{center}

\vspace{0.1cm} 
\begin{center} 
  \emph{$^{1}$ 
  Max-Planck-Institut f\"ur Physik, F\"ohringer Ring 6, \\ 
  80805 M\"unchen, Germany } \\ 
  \vspace{0.25cm} 
  \emph{$^{2}$ Arnold-Sommerfeld-Center for Theoretical Physics, \\ 
  Department f\"ur Physik, Ludwig-Maximilians-Universit\"at  M\"unchen, \\ 
  Theresienstra\ss e 37, 80333 M\"unchen, Germany} \\ 
\end{center}

\vspace{-0.1cm} 
\begin{center} 
  \tt{ 
  akerblom, blumenha, luest, pumuckl@mppmu.mpg.de \\ 
  } 
\end{center}

\vspace{1.5cm} 
\begin{abstract} 
\noindent  
We review some aspects of D-instantons in intersecting D-brane models. In particular, we present applications of the recently proposed instanton calculus to computations of charged matter superpotential couplings and corrections to the gauge kinetic function in the low energy effective action of type IIA orientifolds. As an interesting byway, we also discuss how one-loop corrections to the gauge kinetic function can be deduced from gauge threshold corrections in the type IIA setting.
\end{abstract} 

\thispagestyle{empty} 
\clearpage 
\tableofcontents 

\section{Introduction}
Strictly speaking, the only fully worked out formulation of string theory that we have is perturbative string theory. But since it is abundantly clear that the perturbative scheme cannot possibly be complete, one has always tried to incorporate non-perturbative effects into string theory, essentially relying on intuition from field theory. In contradistinction to compactifications of the heterotic string, instanton corrections to the low-energy physics of intersecting D-brane models have at first been largely ignored. Recent times, however, have seen an embiggened interest in such effects, often motivated by, amongst other things,  the desire to put on a firm stringy basis arguments only made on the level of the effective theory (cf. stability of the landscape, say). The aim of this paper, then, is to review a select number of topics related to the type IIA charged matter field coupling instanton calculus \cite{Blumenhagen:2006xt}, which is a method of dealing systematically with spacetime instantons in intersecting D-brane models.

More specifically, in section \ref{ingred} we discuss, amongst other things, the zero modes which arise from the presence of a single euclidean D2-brane (E2-brane, for short). These zero modes figure as the integration variables of the charged matter field coupling instanton calculus \cite{Blumenhagen:2006xt}. In section \ref{thresh} we present the result that the one-loop corrections to the gauge kinetic function in type IIA orientifolds can be read off from gauge threshold corrections and illustrate this for the example of the $\mathbb{Z}_2\times\mathbb{Z}_2$ toroidal orbifold \cite{Akerblom:2007uc}. By duality, of course, a similar result holds for the type IIB case as well, but, for the sake of brevity, we do not discuss it here. Section \ref{hol} is devoted to explaining how the instanton calculus manages to yield holomorphic contributions to the charged matter field superpotential which is, of course, an important consistency condition \cite{Akerblom:2007uc}. In section \ref{corr} we touch on the question of instanton corrections to the gauge kinetic function, essentially giving necessary conditions for non-vanishing corrections \cite{Akerblom:2007uc}. Finally, in section \ref{further} we cite some topics not discussed in this article.

\section{Zero modes of the charged matter field coupling instanton calculus}\label{ingred}
Intersecting brane models have been the subject of intensive work in the last few years \cite{Uranga:2003pz,MarchesanoBuznego:2003hp,Lust:2004ks,Blumenhagen:2005mu,Blumenhagen:2006ci}. They are given by compactifying
type IIA string theory on an orientifolded six-dimensional compact space which will be assumed to be a smooth
Calabi-Yau space or a singular limit thereof, namely an orbifolded six-torus.
The non-perturbative corrections to be discussed in this article are given by D$p$-branes wrapping non-trivial cycles in the compactification
space and being pointlike in four-dimensional spacetime. Their worldvolume is $(p+1)$-dimensional and spacelike, and thus they will be
referred to as euclidean D$p$-branes, in short E$p$-branes or E$p$-instantons. In type IIA string theory the candidates are then E0-, E2- and E4-branes,
but, as there are no topologically non-trivial one- and five-cycles on a Calabi-Yau, only
E2-branes are relevant and will be discussed in the following.

As already hinted at in the introduction, the prescriptions for determining the effects of such instantons cannot be
derived from first principles but have to be guessed from what is known about field theory instantons. It is therefore plausible
that the number and structure of zero modes is essential. The zero modes are here given by massless open strings with
at least one end on an E2-brane. The following types of zero modes will be distinguished \cite{Blumenhagen:2006xt} (see figure \ref{zero}):

\begin{figure}\label{zero}
\begin{center}
  \includegraphics[width=0.6\textwidth]{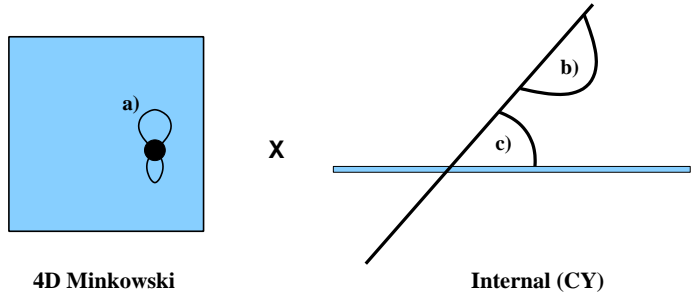}
\end{center}
\caption{Visualization of the origin of the various zero modes. A D6-brane is shown in light blue, while an E2-brane is represented as a black blob in Minkowski space together with a line in the internal space. The E2-brane gives rise to: a) universal zero modes, and possibly b) deformation zero modes as well as c) charged zero modes.}
\end{figure}

\paragraph{Universal zero modes:} These zero modes come from strings starting and ending on the (same) E2-brane. Firstly, there are four bosonic zero modes corresponding to
translational invariance, broken by the instanton, in four-dimensional spacetime and thus parameterising the
position of the instanton. Secondly, there are fermionic zero modes related to broken supersymmetries. By analogy with field theory we require that D-instantons are stationary points of the (fictitious) spacetime ``string action.'' This translates to considering stationary points of the DBI action, which in our case are E2-branes wrapping special Lagrangian 3-cycles in the compactification manifold. The E2-branes
discussed here are thus half-BPS states which implies that they break half of the supercharges.
Two cases have to be
distinguished. Either the submanifold (not only its homology class) wrapped by the instantonic brane is not invariant
under the spacetime action of the orientifold projection. Then the brane ``sees'' the eight supercharges preserved
in the bulk upon compactifying type IIA string theory on a Calabi-Yau and there are four fermionic zero modes.
The gauge symmetry on the instanton in this case is unitary. Or,
the brane is invariant under the orientifold projection. It will be assumed that the orientifold projection is
such that the gauge symmetry on the brane is orthogonal rather than symplectic,\footnote{There are more zero modes if the gauge symmetry
is sympletic. This case will not be discussed in this article.} in which case there are two fermionic
zero modes \cite{Argurio:2007qk,Argurio:2007vqa,Bianchi:2007wy,Ibanez:2007rs}. The other two that would be present in the absence of the orientifold projection are projected out by the latter.

\paragraph{Deformation zero modes:} There can be further zero modes due to possible deformations of the 3-cycle $\Xi$ which the
instanton wraps \cite{Blumenhagen:2006xt}. These are counted by the first Betti number of $\Xi$,  that is, $b_1(\Xi)$.
For an instanton with unitary gauge group there are then $4b_1$ fermionic and $2b_1$ bosonic zero modes. In the case
of an orthogonal (or symplectic) gauge group some of these are projected out and there will be, depending on the
orientifold projection, either $2b_1$ fermionic and $2b_1$ bosonic or $2b_1$ fermionic and no bosonic zero modes. Further zero modes
can arise in the unitary case from the intersection of the brane with the orientifold plane and with its orientifold image.

\paragraph{Charged zero modes:} Finally, in any particular intersecting D6-brane model, there will generically be zero modes which arise at the intersection of the instanton with the D6-branes \cite{Blumenhagen:2006xt}. Denoting the cycle which brane stack $a$ wraps by $\Pi_a$, there will be
$\Pi_a\cap\Xi+\Pi_a\cap\Xi'$ fermionic zero modes associated with this stack. These zero modes are called charged zero modes as they are charged
under the four-dimensional gauge symmetry. Obviously, one has to take into account the intersections of $\Xi$ with all D6-branes.
\bigskip

A particular observable one might want to calculate in an E2-brane background is the (spacetime) correlator of a given number of charged matter fields (cf. also section \ref{hol}). In this case, an important selection rule for determining the possible correlators (i.e. those which have a chance of being non-vanishing) comes from anomalous
$U(1)$ gauge symmetries \cite{Blumenhagen:2006xt}. As a consequence of being anomalous, they are massive and do not appear as gauge symmetries
in the low-energy effective action, but they still persist as global $U(1)$ symmetries in string perturbation theory due to worldsheet boundary combinatorics. Though these symmetries are broken by E2-instantons, they are still symmetries of the low-energy effective action, e.g. the superpotential, in the following sense:

The superpotential
\begin{eqnarray}
 W = \prod_i \Phi_i \exp \left( - S_{\mathrm{E}2} \right)
 \label{superpotex}
\end{eqnarray}
is allowed iff the $U(1)$ gauge transformation of the product of matter fields $\prod_i \Phi_i$ is cancelled
by the transformation of the exponential factor in \eqref{superpotex} induced by a shift of the imaginary part of
the instanton action $S_{\mathrm{E}2}$ under the $U(1)$ symmetry. This imaginary part is the axion field stemming from
integrating the RR three form potential over the 3-cycle the instanton wraps, in other words the Chern-Simons
action of the instanton.

The transformation of the instanton action under the $U(1)$ symmetries can be determined from the
sum of the $U(1)$ charges of the instanton zero modes \cite{Blumenhagen:2006xt}. This is best illustrated in an example.
Consider an anomalous gauge symmetry $U(1)_a$. According to the rules given above, one finds
$\Pi_a\cap\Xi$ instanton zero modes charged under $U(1)_a$. The sign of the charge depends on whether
the intersection number is positive or negative. The sum of the charges of all zero modes is then given by
the topologial intersection number $\Pi_a\circ\Xi$. Taking into account also the orientifold image $\Pi_a'$ of $\Pi_a$, 
one finds that the $U(1)_a$ charge of the instanton is given by
\begin{eqnarray}
 Q_a(\mathrm{E}2) = \mathcal{N}_a \ \Xi \circ \left( \Pi_a - \Pi'_a \right) ,
\end{eqnarray}
where $\mathcal{N}_a$ denotes the number of branes on the stack $a$. This means that the instanton action
transforms under a $U(1)_a$ gauge transformation with transformation parameter $\Lambda_a$ as
\begin{eqnarray}
 \exp \left( - S_{\mathrm{E}2} \right) \rightarrow \exp \left( 2 \pi i\, Q_a(\mathrm{E}2)\, \Lambda_a \right) \exp \left( - S_{\mathrm{E}2} \right) \ .
\end{eqnarray}
Thus, in order to figure out whether a superpotential term is allowed one has to check invariance under all
$U(1)$ symmetries of the model. This selection rule also comes out of the stringy computation of the correlator by means of a complete absorption of all charged fermionic zero modes.

Finally, the holomorphy of the superpotential and the gauge kinetic function together with the string coupling dependence
of the amplitudes and the vertex operators allow for a deduction of which conformal field theory amplitudes are relevant when computing
a (spacetime) correlation function. In the low-energy field theory, the string coupling is given by the VEV of the dilaton, which
is the real part of a chiral supermultiplet, the imaginary part of which enjoys a shift-symmetry. The dependence of the holomorphic
quantities on this multiplet is therefore strongly restricted.

It was argued in \cite{Blumenhagen:2006xt} that the vertex operators for the charged zero modes should carry a factor $g_s^{1/2}$ to ensure that the
instanton decouples for vanishing string coupling. Furthermore, a disc diagram carries a normalisation factor $g_s^{-1}$,
an annulus $g_s^0$ and higher genus amplitudes a positive power of the string coupling. The latter immediately implies that
these higher genus diagrams do not contribute when computing corrections to holomorphic functions.

\section{Threshold corrections and the holomorphic gauge kinetic function}\label{thresh}
In this section we explain how the holomorphy of the gauge kinetic function can be used to infer its one-loop corrections. This involves a result concerning the running coupling in supersymmetric gauge field theories, the so-called Kaplunovsky-Louis formula \cite{Shifman:1986zi,Kaplunovsky:1994fg,Kaplunovsky:1995jw}. As to the organization of this section, we first state a non-renormalization theorem for the gauge kinetic function in type IIA intersecting D-brane models and then review the Kaplunovsky-Louis formula, before putting together these two pieces to derive the connection between gauge threshold corrections and the gauge kinetic function.
\subsection{A non-renormalization theorem for the gauge kinetic function}
We consider a type IIA intersecting D6-brane model with O6-planes on some Calabi-Yau (or toroidal orbifold) $X$. Then the tree-level gauge kinetic function for the gauge theory on brane stack $a$ is given by
\begin{equation}\label{treeresult}
f_a=\sum_I M_{a}^{I}\,U_{I}^{c}\,,
\end{equation}
where the complexified complex structure moduli \( U_{I}^{c} \) are given by integrals along the $A$-cycles of a symplectic basis of homological three-cycles on $X$:
\begin{equation}
U_{I}^{c}=\frac{1}{2\pi\ell_{s}^{3}}\left[e^{-\phi_4}\int_{A_{I}}\mathrm{Re}(\widehat{\Omega}_3)-i\int_{A_{I}}C_3\right]\,.
\end{equation}
Using holomorphy and the Peccei-Quinn shift-symmetry of the complexified complex structure moduli, one infers that the full gauge kinetic function (including the tree level result and all corrections) will have the functional form
\begin{equation}\label{funcform}
f_a=\sum_{I}M_{a}^I\,U_{I}^{c}+f_{a}^{\text{1-loop}}(e^{-T_{i}^{c}})+f^{\text{non pert.}}(e^{-U_{I}^{c}},e^{-T_{i}^{c}})\,,
\end{equation}
where the complexified K\"ahler moduli \( T_{i}^{c} \) are defined by integrals along a basis of two-cycles \( C_i \in H_{-}^{1,1}(X) \) anti-invariant under the orientifold projecction:
\begin{equation}
T_{i}^{c}=\frac{1}{\ell^{2}}\left(\int_{C_{i}}J_2-i\int_{C_{i}}B_2\right)\,.
\end{equation}
We would like to draw the reader's attention to the fact that the form of \eqref{funcform} is vindicated in a number of specific examples \cite{Akerblom:2007np,Camara:2007dy,Blumenhagen:2007ip}.
\subsection{Gauge threshold corrections and one-loop corrections to the gauge kinetic function}
If we consider a supersymmetric gauge field theory induced on a stack of branes, there are two ways to determine its running coupling.

The first is via the standard equation for the running coupling
\begin{equation}\label{running}
\frac{8\pi^{2}}{g_a^2(\mu)}=\frac{8\pi^{2}}{g_{a, \text{string}}^2}+\frac{b_a}{2}\,\log\left(\frac{M_s^2}{\mu^2}\right)+\frac{\Delta_a}{2}\, ,
\end{equation}
where $\Delta_a$ denotes the gauge threshold corrections, coming from integrating out massive particles running in loops and $b_a$ is the one-loop beta-function coefficient.\footnote{As before, the index $a$ labels the particular brane-stack one is interested in.} In general, the gauge threshold corrections are the sum of two terms: one depends non-holomorphically on the closed string moduli, whereas the second is the real part of a holomorphic function
\begin{equation}\label{sumhol}
\Delta_a=\Delta_a^{\text{n.h.}}+\mathrm{Re}(\Delta_a^{\text{hol.}})\,.
\end{equation}

The second way of calculating the running coupling is via the Kaplunovsky-Louis formula, applied to the situation at hand \cite{Shifman:1986zi,Kaplunovsky:1994fg,Kaplunovsky:1995jw,Derendinger:1991kr,LopesCardoso:1991zt,Derendinger:1991hq,Ibanez:1992hc}:
\begin{multline}\label{KLformula}
\frac{8\pi^2}{g_a^2(\mu^2)}=8\pi^2\,\mathrm{Re}(f_a)+\frac{b_a}{2}\, \log\left(\frac{M_s^2}{\mu^2}\right)+\frac{c_a}{2}\, \mathcal{K}+
\\+T(G_a)\,\log\left(g_a^{-2}(\mu^2)\right)-\sum_rT_a(r)\,\log\left(\det K_r(\mu^2)\right)\, ,
\end{multline}
with
\begin{equation*}
b_a=\sum_r n_r\,T_a(r)-3T(G_a) \quad \text{and} \quad c_a=\sum_r n_r\,T_a(r)-T(G_a)\,.
\end{equation*}
In this formula, $G_a$ is the gauge group of the field theory induced on brane stack $a$, the $T_{(a)}$ are the generators of the group and $T_a(r)=\mathrm{Tr}(T^2_{(a)})$, while $\mathcal{K}$ denotes the tree-level closed string moduli K\"ahler potential and $K$ the tree-level matter K\"ahler metric. Furthermore, $r$ runs over the representations of $G_a$, with $n_r$ the number of multiplets in the representation (labeled by) $r$ and $T(G_a)=T_a(\mathrm{adj}\,G_a)$. The gauge kinetic function $f_a$ on the right of \eqref{KLformula} is understood to be meant in the Wilsonian sense, which implies that it is holomorphic.

Equating the right hand sides of \eqref{running} and \eqref{KLformula} we then deduce with the help of \eqref{treeresult}, \eqref{funcform} and \eqref{sumhol} that
\begin{equation}
f_a=f_a^{\text{tree}}+f_a^{\text{1-loop}}(e^{-T_i^c})=\sum_I M_a^I U_I^c+\Delta_a^{\text{hol.}}\, ,
\end{equation}
from which it follows that
\begin{equation}\label{threshrel}
f_a^{\text{1-loop}}=\Delta_a^{\text{hol.}}\,.
\end{equation}

\subsection{Example: The $T^6/\mathbb{Z}_2\times\mathbb{Z}_2$ orbifold}

In the case of the type IIA $T^6/\mathbb{Z}_2\times\mathbb{Z}_2$ orientifolded orbifold, the gauge threshold corrections are known explicitly \cite{Lust:2003ky,Akerblom:2007np}.\footnote{Recently, also the case of the $T^6/\mathbb{Z}_2\times\mathbb{Z}'_2$ orbifold has been worked out \cite{Blumenhagen:2007ip}.} For the sake of brevity, we cite these results not in full generality but impose some restrictions on the intersection angles of the branes.

In a sector preserving $\mathcal{N}=1$ supersymmetry the gauge threshold corrections are
\begin{equation}
\Delta_a^{\mathcal{N}=1}=-\frac{b_a}{16\pi^2}\log\left[\frac{\Gamma(\theta^1_{ab})\Gamma(\theta^2_{ab})\Gamma(1+\theta^3_{ab})}{\Gamma(1-\theta^1_{ab})\Gamma(1-\theta^2_{ab})\Gamma(-\theta^3_{ab})}\right]\, , \quad \text{for} \quad \theta^{1,2}_{ab}>0, \, \theta^3_{ab}<0\, .
\end{equation}
Since the $\theta$'s are non-holomorphic functions of the closed string moduli, it follows that this has no holomorphic piece. Thus we have
\begin{equation}
\mathcal{N}=1: \quad f_a^{\text{1-loop}}=0\,.
\end{equation}

In a sector preserving $\mathcal{N}=2$ supersymmetry, one deduces from the annulus amplitude
\begin{equation}
T^A(\mathrm{D6}_a,\mathrm{D6}_b)\propto\log\left(\frac{M_s^2}{\mu^2}\right)-\log \left|\eta(i T^c_I)\right|^4-\log(\mathrm{Re}(T^c_I)\, V_I^a)+\gamma_E-\log4\pi\, ,
\end{equation}
with $V_I^a$ the volume of the 1-cycle that brane stack $a$ wraps on the $I$'th 2-torus, that
\begin{equation}
\mathcal{N}=2:\quad f_a^{\text{1-loop}}\propto \log(\eta(iT_I^c))\, .
\end{equation}
\section{Holomorphy of the instanton-induced superpotential}\label{hol}
In the following, the conditions for an E2-instanton to contribute to the superpotential will be discussed and it will be shown
that, although not clear a priori, the instanton-generated couplings can be incorporated in a (holomorphic) superpotential.
The instanton measure must contain a factor $d^4x\,d^2\theta$ in order for a superpotential contribution to arise, and there must
not be any further uncharged zero modes \cite{Blumenhagen:2006xt}. From the discussion of the zero modes in section \ref{ingred} it is clear that
there must not be any deformation zero modes, in other words the three-cycle wrapped by the instanton must be rigid. Furthermore,
one sees that the zero mode structure of an instanton with orthogonal gauge group is appropriate for a superpotential to arise,
whereas an instanton with unitary gauge group at first sight has two zero modes too many \cite{Argurio:2007qk,Argurio:2007vqa,Bianchi:2007wy,Ibanez:2007rs}. However, these ``superfluous'' zero modes can be absorbed if the
instanton wraps a cycle that is also wrapped by a D6-brane, which means that the instanton is a gauge instanton. The integration over these two zero modes then produces the
fermionic ADHM constraints. In this section only instantons with orthogonal gauge group will be considered.

The discussion in section \ref{ingred} of the interplay between the string coupling dependence of certain amplitudes and the holomorphy
of certain quantities implies for the superpotential to be considered in this section that, apart from
the vacuum disc diagram yielding the exponential of the instanton action, only disc diagrams with precisely two
charged zero modes inserted and annulus diagrams (without charged zero modes inserted) contribute \cite{Dorey:2002ik,Billo:2002hm,Blumenhagen:2006xt}.

\begin{figure}[t]
\begin{center}
  \includegraphics[width=0.7\textwidth]{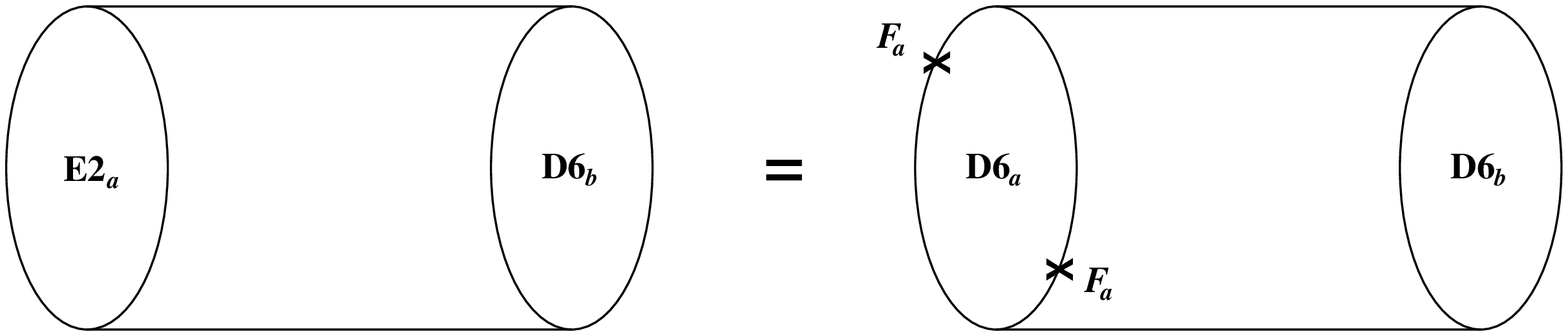}
\end{center}
\caption{Relation between instantonic one-loop amplitudes and 
corresponding
  gauge  threshold corrections.}\label{figureINSTTH}
\end{figure}

A (spacetime) correlator of charged matter fields in the instanton background is computed as follows \cite{Blumenhagen:2006xt}:
\begin{eqnarray}
      &&\hskip -1cm \langle \Phi_{a_1,b_1}\cdot\ldots\cdot   \Phi_{a_M,b_M}  
\rangle_{\mathrm{E}2}  
 = \frac{V_3}{g_s}\,\int d^4 x\, d^2\theta \,\, 
       \sum_{\rm conf.}\,\,  {\textstyle  
  \prod_{a} \bigl(\prod_{i=1}^{ [\Xi\cap 
             \Pi_a]^+}  d\lambda_a^i\bigr)\, 
               \bigl( \prod_{i=1}^{ [\Xi\cap 
             \Pi_a]^-}  d\overline{\lambda}_a^i\bigr) } \nonumber \\ 
   && \phantom{a} \exp ({-S_{\mathrm{E}2}}) \,\, 
           \exp \left({Z'_0(\mathrm{E}2)}\right) \,\, 
\langle \widehat\Phi_{a_1,b_1}[\vec x_1]   \rangle_{\lambda_{a_1},\overline{\lambda}_{b_1}}\cdot 
            \ldots \cdot  \langle \widehat\Phi_{a_L,b_L}[\vec x_L] 
          \rangle_{\lambda_{a_L},\overline{\lambda}_{b_L}}\,.
 \label{spacetimecorrelator}
\end{eqnarray}
It involves an integration over all instanton zero modes and a sum over all configurations of distributing the vertex operators
for the charged matter fields $\Phi_{a_i,b_i}$ on disc diagrams, on each of which two charged zero modes are
inserted.\footnote{Note that for simplicity the possibility of including annuli with charged matter
fields inserted is neglected.} The abbreviation $\widehat\Phi_{a_k,b_k}[{\vec x_k}]$ denotes
a chain-product of vertex operators
\begin{eqnarray}
  \widehat\Phi_{a_k,b_k}[{\vec x_k}] = \Phi_{a_k,x_{k,1}}\cdot  
  \Phi_{x_{k,1},x_{k,2}} \cdot \Phi_{x_{k,2},x_{k,3}}\cdot \ldots 
  \cdot \Phi_{x_{k,n-1},x_{k,n}}  \cdot \Phi_{x_{k,n(k)},b_k} \quad ,
\end{eqnarray}
while 
$\langle \widehat\Phi_{a_1,b_1}[\vec x_1]   \rangle_{\lambda_{a_1},\overline{\lambda}_{b_1}}$ is a CFT disc correlator
with the vertex operators for $\widehat\Phi_{a_1,b_1}[\vec x_1]$ and those for the charged zero modes
$\lambda_{a_1}$ and $\overline{\lambda}_{b_1}$ inserted on the boundary. Finally, $V_3$ is the volume
of the three cycle the instanton wraps and $Z'_0(E2)$ is the sum of all 
annulus vacuum amplitudes in the instanton background
\begin{eqnarray} 
   \langle 1 \rangle^{\text{1-loop}}=   
    Z'_0(\mathrm{E}2) = 
   {\textstyle \sum_b  {Z'}^A ({\rm E2}_a,{\rm D6}_b)  
    +  {Z'}^M({\rm E2}_a,{\rm O6})}\; ,
   \label{fulloneloopamplitude}
\end{eqnarray}
where the prime \cite{Blumenhagen:2006xt} denotes omission of the zero modes. They are omitted as they are integrated over explicitly in
the correlator.
The one-loop diagrams are
\begin{eqnarray}
    Z^A({\rm E2}_a,{\rm D6}_b)&=&-\int_0^\infty  \frac{dt}{t}\,
   \sum_{\alpha,\beta \neq (\frac{1}{2},\frac{1}{2})} (-1)^{2(\alpha+\beta)}\,
    \frac{ \vartheta^2\genfrac[]{0pt}{}{\alpha}{\beta}(it/2,it) }{
      \vartheta^2\genfrac[]{0pt}{}{1/2}{1/2}(it/2,it)}\, \frac{\eta^3(it) }{
\vartheta\genfrac[]{0pt}{}{\alpha}{\beta}(0,it)}
    \,A^{\rm CY}_{ab}\genfrac[]{0pt}{}{\alpha}{\beta}\nonumber
\\ &=& -\frac{1}{4\pi^2}\int_0^\infty  \frac{dt}{t}\,
   \sum_{\alpha,\beta\neq (\frac{1}{2},\frac{1}{2})} (-1)^{2(\alpha+\beta)}\,
   \frac{\partial^2_\nu\vartheta\genfrac[]{0pt}{}{\alpha}{\beta}(\nu,it)|_{\nu=0} }{\eta^3(it)}\,\,
   A^{\rm CY}_{ab} \genfrac[]{0pt}{}{\alpha}{\beta} \;.
 \label{oneloopamplitude}
\end{eqnarray}
From the latter form in \eqref{oneloopamplitude} one can see that the one-loop amplitudes are equal to the
gauge threshold corrections discussed in the previous section \cite{Abel:2006yk,Akerblom:2006hx}. This equality is
diagrammatically shown in figure \ref{figureINSTTH}.
A similar relation holds for the M\"obius diagrams. As the threshold corrections are not holomorphic functions
of the closed string moduli, it is not clear whether the correlator \eqref{spacetimecorrelator} depends holomorphically 
on the moduli and therefore whether the coupling which it describes can be incorporated in a superpotential. Using the
Kaplunovsky-Louis formula already employed in section \ref{thresh}, it will now be shown that this is actually the case.

Applying this formula to the instanton amplitude \eqref{fulloneloopamplitude}, where the single contributions
are given by \eqref{oneloopamplitude}, one finds that
\begin{multline}
Z_0(\mathrm{E}2_a)=   
-8\pi^2\,\mathrm{Re} ( {f^{(1)}_a}) - \frac{b_a}{2}\log\left( \frac{M_{p}^2}{ 
    \mu^2}\right) - \frac{c_a}{2}\, {\cal K}_{\rm tree}\\
  -\log\left( \frac{V_3}{g_s}\right)_{\rm tree} + \sum_b \frac{|I_{ab} N_b|}{2} 
\log\left[\det K^{ab} \right]_{\rm tree}\, , \label{KLinstanton}
\end{multline}
where for the brane and instanton configuration in question the coefficients 
are   
\begin{eqnarray}
   b_a=\sum_b \frac{|I_{ab} N_b|}{2} -3,\quad\quad  
   c_a=\sum_b \frac{|I_{ab} N_b|}{2} -1. 
\end{eqnarray}
To proceed, note that in string theory, physical amplitudes are calculated, which means that the CFT amplitudes
in \eqref{spacetimecorrelator} involve holomorphic couplings $Y_{\lambda_a\, \widehat\Phi_{a,b}[x]\, \overline\lambda_b}$
and non-holomorphic K\"ahler potential/metric terms in the usual way \cite{Cremades:2003qj,Cvetic:2003ch,Abel:2003vv,Cremades:2004wa}
\begin{eqnarray}
\langle \widehat\Phi_{a,b}[\vec x]   \rangle_{\lambda_{a},\overline{\lambda}_{b}}&=& 
\frac{e^\frac{{\cal K}}{ 2}\, Y_{\lambda_a \Phi_{a,x_1}\Phi_{x_1,x_2}\ldots 
   \Phi_{x_N,b}\,  \overline\lambda_b}  
    }{ \sqrt{ K_{\lambda_a,a} \, K_{a,x_1}\, \ldots  K_{x_n,b}\,  
          K_{b,\overline\lambda_b} }} 
   = \frac{
e^\frac{{\cal K}}{2}\, Y_{\lambda_a\, \widehat\Phi_{a,b}[x]\, \overline\lambda_b}  
    }{ \sqrt{ K_{\lambda_a,a} \, \widehat K_{a,b}[x]\,  
          K_{b,\overline\lambda_b} } } \, . \label{CFTholokaehler}
\end{eqnarray}
Analogously, the spacetime correlator \eqref{spacetimecorrelator} should
involve the superpotential Yukawa couplings as well as terms involving the K\"ahler potential/metric
\begin{eqnarray}
      && \langle \Phi_{a_1,b_1}\cdot\ldots\cdot   \Phi_{a_M,b_M}  
\rangle_{\mathrm{E}2\text{-inst}} =  
\frac{e^\frac{{\cal K}}{2}\, Y_{\Phi_{a_1,b_1},\ldots, \Phi_{a_M,b_M}} }{ \sqrt{ K_{a_1,b_1}\cdot\ldots\cdot K_{a_M,b_M}   }}. 
 \label{Yukawaholokaehler}
\end{eqnarray}
Inserting \eqref{KLinstanton} and \eqref{CFTholokaehler} into \eqref{spacetimecorrelator} one finds that the non-holomorphic
K\"ahler terms rearrange in such a way that \eqref{spacetimecorrelator} takes the form \eqref{Yukawaholokaehler}
with the superpotential Yukawas being holomorphic and given by
\begin{multline}
     Y_{\Phi_{a_1,b_1},\ldots, \Phi_{a_M,b_M}}= 
       \sum_{\rm conf.}\,\,  \sign_{\rm conf.}\, 
  \,\, \exp ({-S_{\mathrm{E}2}})_{\rm tree}\cdot\\ 
        \cdot \exp \left(-f^{(1)}_a\right)  
Y_{\lambda_{a_1}\, \widehat\Phi_{a_1,b_1}[\vec x_1]\, \overline\lambda_{b_1}} 
\cdot\ldots\cdot Y_{\lambda_{a_1}\, \widehat\Phi_{a_L,b_L}[\vec x_L]\, 
         \overline\lambda_{b_L}} .
 \label{Yukawaholo}
\end{multline}
Note here that for the term $\exp({\cal K}/2)$ required in \eqref{Yukawaholokaehler} to come out correctly, it is crucial
that precisely two charged fermion zero modes appear in each disc correlator, confirming this conjecture. Note furthermore,
that higher genus amplitudes can contribute to the correlator \eqref{spacetimecorrelator}, but they are related to
$g_s$-corrections to the K\"ahler metric/potential and \eqref{Yukawaholo} is exact.

\section{Instanton corrections to the gauge kinetic function}\label{corr}
In section \ref{thresh} we discussed one-loop corrections to the gauge kinetic function, explaining how they could be inferred from the knowledge of gauge threshold corrections, thereby allowing us to determine the \emph{full perturbative} gauge kinetic function for the $\mathcal{N}=1$ and $\mathcal{N}=2$ sectors of the $T^6/\mathbb{Z}_2\times\mathbb{Z}_2$ type IIA orbifold. Further corrections are necessarily non-perturbative, so let us now consider under which conditions such corrections can arise.

By applying S- and T-dualities to the story of world-sheet instanton corrections in the heterotic string, we indeed expect that there can be E2-instanton corrections to the gauge kinetic function. In the heterotic case, similar to the topological type IIA string, such corrections arise from string world-sheets of Euler characteristic zero. In the (dual) case at hand they are therefore expected to arise from world-sheets with two boundaries. Also, as an additional condition, we expect such corrections to appear for E2-instantons admitting one complex open string modulus, i.e. for those wrapping a 3-cycle $\Xi$ with Betti number $b_1(\Xi)=1$.

For the following let us recall the instanton zero mode structure 
for such a cycle (cf. ``deformation zero modes'' in section \ref{ingred}): Before the orientifold projection, each (complex) bosonic field (zero mode) $y$ is accompanied by two pairs of fermionic zero modes $\mu_\alpha$ and $\ov{\mu}_{\dot{\alpha}}$.

\begin{figure}
\begin{center} 
 \includegraphics[width=0.38\textwidth]{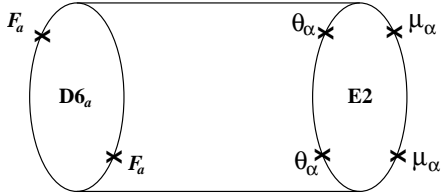} 
\end{center} 
\caption{Annulus diagram for E2-instanton correction to $f_a$.}\label{finst} 
\end{figure}

For the orientifolding, one has to distinguish two cases according to the two distinct ways the   
anti-holomorphic involution $\ov\sigma$ can act on the open string 
modulus $y$: 
\begin{equation} 
      \ov\sigma:y\to \pm y\,. 
\end{equation} 
In the case that $y$ is invariant under $\ov\sigma$, called first kind in the 
following,  the orientifold 
projection acts in the same way as for the 4D fields $X^\mu$ and $\theta_\alpha$, i.e. 
the (complex) bosonic zero mode $y$\footnote{Note that there are thus two real bosonic degrees of freedom.} and the two fermionic zero modes 
$\ov\mu_{\dot{\alpha}}$ survive. In the other case, which we call second kind,  the bosonic zero mode is projected 
out and only the fermionic modulino zero modes $\mu_\alpha$ survive.\footnote{By duality, this 
distinction is related to the two kinds of deformations 
of genus $g$ curves studied in \cite{Beasley:2005iu}. The first kind corresponds to 
curves moving in families, i.e. to transversal deformations of the curve, while 
the second kind is related to  
those deformations coming with the genus $g$ of the curve.} 
Therefore, in the absence of any additional zero modes, for instance 
from E2-D6 intersections, the zero mode measure in any instanton 
amplitude assumes the following 
form 
\begin{equation} 
      \int d^4 x\, d^2\theta\, d^2 y\, d^2 \ov\mu\ e^{-S_{E2}}\ldots,\quad\quad {\rm for}\  
         \ov\sigma:y\to y 
\end{equation} 
and 
\begin{equation} 
      \int d^4 x\, d^2\theta\, d^2 \mu \ e^{-S_{E2}}\ldots,\quad\quad {\rm for}\  
         \ov\sigma:y\to -y\,. 
\end{equation} 
It is clear that only an instanton with precisely one set of  fermionic zero 
modes of the second kind and no additional zero modes 
can generate a correction to the $SU(N_a)$ gauge kinetic function. The  
amplitude in the instanton background takes the form
\begin{equation} 
\label{gaugekin} 
     \langle F_a(p_1)\, F_a(p_2)  
\rangle_{\mathrm{E}2}  
 = \int d^4 x\, d^2\theta \, d^2 \mu\ 
       \exp ({-S_{\mathrm{E}2}}) 
         \, \exp \left({Z'_0 (\mathrm{E}2)}\right)\,  A_{F_a^2}(\mathrm{E}2,\mathrm{D}6_a) \,,
\end{equation} 
where $A_{F_a^2}(\mathrm{E}2,\mathrm{D}6_a)$ is the annulus diagram in figure \ref{finst}, in which 
all the appearing fermionic zero modes are absorbed. 

\section{Further developments}\label{further}
Before closing, let us mention a few recent developments related in one way or another to the topics of this article. For reasons of space, our selection will necessarily be quasi-random.

The observation that one-loop instanton amplitudes are equal to gauge threshold corrections and the fact that they  depend non-holomorphically on the closed string moduli has triggered some interest in gauge threshold corrections and their relation to the one-loop correction to the gauge kinetic function, as discussed in  section \ref{thresh}. Using the Kaplunovsky-Louis formula for this relation also constrains the form of the chiral matter K\"ahler metric \cite{Akerblom:2007uc,Billo:2007sw,Billo:2007py,Blumenhagen:2007ip}.

A lot of work has been concerned with the derivation of instanton-induced superpotential contributions \cite{Blumenhagen:2006xt,Haack:2006cy,Ibanez:2006da,Abel:2006yk,Florea:2006si,Akerblom:2006hx,Bianchi:2007fx,Cvetic:2007ku,Argurio:2007vqa,Bianchi:2007wy,Ibanez:2007rs,Blumenhagen:2007zk,Cvetic:2007qj}, on the one hand reproducing Affleck-Dine-Seiberg type superpotentials, on the other hand determining terms induced by stringy instantons. Concerning the latter case, the main focus has been on neutrino Majorana masses \cite{Ibanez:2006da,Cvetic:2007ku,Ibanez:2007rs,Cvetic:2007qj}.

One important result is that for a stringy instanton to yield a contribution to the superpotential, the gauge group on its world volume must be orthogonal \cite{Argurio:2007qk,Argurio:2007vqa,Bianchi:2007wy,Ibanez:2007rs}, though it has been realized recently that under certain circumstances this requirement can be evaded \cite{Petersson:2007sc}. Furthermore, instanton recombination and background fluxes have been discussed as a way to get rid of some of the zero modes \cite{Blumenhagen:2007bn}.

Multi-instanton effects and issues related to changes in the instanton structure at certain loci in the moduli space have just started to attract attention \cite{Ibanez:2007tu,GarciaEtxebarria:2007zv}.

Finally, duality arguments have allowed to determine non-perturbative corrections to the gauge kinetic functions \cite{Camara:2007dy}, in agreement with the non-renormalization theorem of section \ref{thresh}.

\vskip 1cm 
 {\noindent  {\Large \bf Acknowledgements}} 
 \vskip 0.5cm
It is a pleasure to thank Sebastian Moster, Erik Plauschinn and Stephan Stieberger for helpful discussions. Erik Plauschinn also helped with the preparation of the figures, for which we are grateful to him.

\nocite{*} 
\bibliography{rev} 
\bibliographystyle{utphys}
\end{document}